\documentclass[useAMS]{mn2e}
\usepackage{epsf}
\usepackage{german}
\usepackage[dvips]{graphics}

\begin{document}
\title[Testing for primordial
non-Gaussianity using surrogates]
{Analysing large scale structure: II. Testing for primordial
non-Gaussianity in CMB maps using surrogates}

\author[C. R\"ath and  P. Schuecker]
  {Christoph R\"ath \thanks{E-mail: cwr@mpe.mpg.de} and
    Peter Schuecker\\
    Centre for Interdisciplinary Plasma Sciences (CIPS)/ \\
    Max-Planck-Institut f\"ur extraterrestrische Physik (MPE),
    Garching, Germany
  }

\date{Accepted ...
      Received ...;
      in original form ...}

%\pubyear{2002} \volume{000} \pagerange{1} \onecolumn
\pubyear{2002} \volume{000} \pagerange{1} \twocolumn

\maketitle

\label{firstpage}

\begin{abstract}

The identification of non-Gaussian signatures in cosmic
microwave background (CMB) temperature maps  is one of
the main cosmological challenges today. We propose and investigate
altenative methods to analyse CMB maps.
Using the technique of constrained randomisation
we construct surrogate maps which mimic both the power spectrum
{\it and} the amplitude distribution of simulated CMB maps
containing non-Gaussian signals.
Analysing the maps with weighted scaling indices and Minkowski functionals
yield in both cases statistically significant identification of the primordial
non-Gaussianities. We demonstrate that the method is
very robust with respect to noise. We also show that Minkowski functionals are
able to account for non-linearities at higher noise level when applied in combination
with surrogates than when only applied to noise added CMB maps and phase randomised
versions of them, which only reproduce the power spectrum.

\end{abstract}

\section{Introduction}

A fundamental question of cosmology is whether
the observed fluctuations were Gaussian or non-Gaussian
when they were formed.
This gives important information about the nature of the fluctuations
and how they were generated (e.g. inflationary or ekpyrotic/cyclic processes,
moving cosmic strings etc.).
A measurement of non-Gaussianity from the large scale structure of galaxies
is, however, difficult because the non-linear gravitational collapse
itself generates non-Gaussianity from Gaussian signals. Therefore,
measurements of the temperature anisotropies of the cosmic microwave
background (CMB) are regarded as the best way to identify the true nature
of primordial fluctuations.
Fundamental limitations of these measurements are cosmic variance on
large scales and the central limit theorem on small scales.\\
In single field inflationary models (Guth 1981; Linde 1982;
Albrecht \& Steinhardt 1982),
incorporating cold dark matter, the distribution of temperature
fluctuations in the CMB should be a homogenous isotropic
almost Gaussian random field.
While for Big-Bang-inspired inflationary scenarios
non-Gaussian contributions of $10^{-5}$ relative to the
leading Gaussian term are expected, for M-theory based
ekpyrotic/cyclic scenarios  non-Gaussian contributions of
higher order are expected to be
exponentially suppressed yielding a kind of super-Gaussianity (Steinhardt \&
Turok 2002).
Multi field inflationary models, on the other hand,
open the door to the generation of primordial non-Gaussianities
because of the possible non-linear couplings of the fields (Bernardeau \& Uzan 2002).
Another class of theories predict the formation
of topological defects such as cosmic strings,
monopoles or textures. According to these theories
the CMB temperature fluctuations are expected to be
non-Gaussian possessing steep gradients or 'hot-spots'
of emission (Bouchet, Bennett \& Stebbins 1988; Turok 1996).\\
The need for very powerful statistical measures for detecting
non-Gaussian signatures in the CMB maps is obvious and has led to
the development of many different analysing techniques.
The bispectrum (e.g. Verde et al. 2000; Sandvik \& Magueijo 2001) and
trispectrum (e.g. Verde \& Heavens 2001; Hu 2001) as well as phase
mapping techniques (Chiang, Coles \& Naselsky 2002;
Chiang, Naselsky  \& Coles 2002) rely on a Fourier-based analysis of
CMB maps.\\
The application of {\it local} Fourier techniques in the form of wavelets
turned out to detect non-Gaussianities with high significance
(Hobson, Jones \& Lasenby 1999; Cay\'{o}n et al. 2001;
Barreiro \& Hobson 2001).\\
Non-Gaussian signatures have also been identified
by quantifying the morphology of CMB maps. In this context
it has been proposed by Coles \& Barrow (1987) and
Coles (1988) to calculate
the genus (Euler characteristic) of an excursion set.
Other measurable topological  quantities are the volume
and circumference of the excursion set.
These three measures can be placed in the wider framework
of Minkowski functionals by natural mathematical
considerations (Mecke, Buchert \& Wagner 1994)
and have found a wide application in the analysis
of  CMB maps (e.g. Smoot et al. 1994;
Kogut et al. 1996;
Schmalzing et G\'{o}rski 1998;
Shandarin et al. 2002).\\
The multifractal formalism has been applied to CMB maps
by Pompillo et al. (1995) and Diego et al. (1999).
The multifractal formalism, however, is not the only
method derived from nonlinear sciences which can
successfully be applied to CMB investigations.
Being aware that it is often very difficult to
doubtlessly identify (multi-)fractal dimensions
for a limited number of points,
the method of surrogates (Theiler et al. 1992, for a review see
Schreiber \& Schmitz 2000)
has been established
in the last years in order to detect (weak)
nonlinearities in time series.
The basic idea is to compute nonlinear statistical measures
for the original data set and of an ensemble of surrogate data sets,
which mimic the linear properties of the original data set.
If the computed measure for the
original data is significantly different from the values obtained
for the surrogate sets, one can infer that the data were generated
by a nonlinear process.\\
Several approaches for the generation of surrogates have been established
depending on the data and on the constraints to be preserved.
Some algorithms make explicit use of Fourier transformations like the
amplitude adjusted Fourier transform algorithm (AAFT) (Theiler et al. 1992)
and the iterative amplitude adjusted Fourier transform
algorithm (ITAAFT) (Schreiber \& Schmitz 1996).
Other approaches use simulated annealing techniques, where
the constraints are implemented as a suitable cost function, which
is to be minimized (Schreiber 1998).\\
So far, these methods have only been applied to time series
analysis, although the general approach is not resticted to
time series analysis.
Recently it has been demonstrated (R\"ath et al. 2002) in the large
scale structure analysis that surrogates can
successfully be generated for three-dimensional point distributions, for
which it is known that they have nonlinear correlations.
It has further been shown that statistical measures which
estimate the local scaling properties of the point set are well
suited to account for the nonlinearities in the data.\\
In this paper we show the feasibility to generate surrogates,
which mimic the power spectrum and the amplitude distribution,
for CMB maps using a two-dimensional version of the
well-known  ITAAFT approach.
As statistical measures for testing for non-Gaussian signatures
in the maps we use weighted scaling indices
as well as Minkowski functionals. 
The results of our studies are compared with those
obtained with established methods. 
In this study we are only interested in a relative comparison
between the different techniques for testing for
non-Gaussian signatures. Therefore we do not consider
effects of cosmic variance because they were also not
taken into account in the studies we compare our results with.\\
The outline of the paper is as follows. Section 2
reviews in some details the method to generate
surrogates for the two-dimensional case.
In Section 3 we introduce as test statistics
weighted scaling indices and the Minkowski
functionals.
The results of applying our method to simulated CMB maps
are shown in Section 4.
Finally, we present our conclusions and give an 
outlook on future studies in Section 5.

%%%%%%%%%%%%%%%%%%%%%%%%%%%%%%%%%%%%%%%%%%%%%%%%%%%%%%%%%%%%%%%%%%%%%
%%%%%%%%%%%%%%%%%%%%%%%%%%%%%%%%%%%%%%%%%%%%%%%%%%%%%%%%%%%%%%%%%%%%%

\section{Surrogates}

An iteration scheme is used to
generate surrogate data which have the same
power spectum and the same temperature  distribution.
As a brief review and to introduce the notations
we describe the extension of the algorithm
for the two-dimensional case of CMB temperature maps.
Assume a two-dimensional pixelized field of random fluctuations in the
CMB brightness temperature $T(x,y) = T(\vec{r}), x,y = 1,\ldots, N $.\\
Before the iteration starts
two quantities have to be calculated:\\
1) A copy $\eta(\vec{r}) = \mbox{rank}(T(\vec{r}))$ of the original
temperature values,
which is sorted
by magnitude in ascending order, is computed.\\
2) The absolute values of the amplitudes of the discrete Fourier
transform $T(\vec{k})$ of  $T(\vec{r})$,
\begin{equation}
   \left| T(\vec{k}) \right| =
   \left| \frac{1}{N^2} \sum_{x,y}
           T(\vec{r}) e^{-2 \pi i \vec{k} \cdot  \vec{r}/N} \right|
\end{equation}
are calculated as well.\\
The starting point for the iteration is a random
shuffle $T_0(\vec{r})$ of the data. Each iteration
consists of two consecutive calculations:\\
First, $T_0(\vec{r})$ is brought to the desired sample
power spectrum. This is achieved by using a crude 'filter' in the
Fourier domain: The Fourier amplitudes are simply {\it replaced} by the
desired ones.\\
For this the Fourier transform of $T_n(\vec{r})$ is taken:
\begin{equation}
 T_n(\vec{k}) = \frac{1}{N^2} \sum_{x,y}
           T_n(\vec{r}) e^{-2 \pi i \vec{k} \cdot \vec{r}/N} \;.
\end{equation}
In the inverse Fourier transformation the actual amplitudes are
replaced by the desired ones and the phases defined
by $ \tan{\psi_n(\vec{k})} = Im(T_n(\vec{k}))/Re(T_n(\vec{k}))$
are kept:
\begin{equation}
 s(\vec{r}) = \frac{1}{N^2} \sum_{k_x,k_y}
           e^{i \psi_n(\vec{k})} \left| T(\vec{k}) \right|
	   e^{-2 \pi i \vec{k} \cdot \vec{r}/N} \;.
\end{equation}
Thus this step enforces the correct power spectrum, but usually the
distribution of the amplitudes in the CMB image will be modified.\\
Second, a rank ordering of the resulting data set $s(\vec{r})$ is
performed in order to adjust the spectrum of amplitudes. The intensities
$T_{n+1}(\vec{r})$ are obtained by replacing the values of
$s(\vec{r})$ with those stored in $\eta(\vec{r})$ according to
their rank:
\begin{equation}
 T_{n+1}(\vec{r}) = \eta(\mbox{rank}(s(\vec{r}))) \;.
\end{equation}
It can heuristically be understood that the iteration scheme
is attracted to a fixed point (for an explanation see e.g.
Schreiber \& Schmitz 2000). The final accuracy that can be reached
depends on the size and properties of the data sets but is generally
sufficient for hypothesis testing.

\section{Test statistics}

\subsection{Weighted scaling indices and their local means}

We use weighted scaling indices (R\"ath et al. 2002)
for the  estimation of {\it local} scaling properties of a
point set and apply
this method in order to characterize different
structural features of the spatial patterns in the images.
Consider a temperature map $T(x,y)$ of size $ M_1 \times M_2$.
Each pixel is assigned with a temperature value $T(x,y)$
thus containing both space and temperature information
that can be encompossed in a three-dimensional vector
$\vec{p} = (x,y,T(x,y))$. The CMB image can now be regarded
as a  set of $N$ points $P=\{\vec{p_i}\}, i=1,\ldots,N, \;, N = | M_1 \times M_2 |$.
For each point the local weighted cumulative point
distribution $\rho$ is calculated.
In general form this can be written as
\begin{equation}
  \rho(\vec{p_i},r) = \sum_{j=1}^{N} s_r (d(\vec{p_i},\vec{p_j})) \;,
\end{equation}
where $s_r(\bullet)$ denotes a kernel function depending on
the scale parameter $r$ and $d(\bullet)$ a distance measure.\\
The weighted scaling indices $\alpha(\vec{p_i},r)$ are obtained by calculating
the logarithmic derivative of $\rho(\vec{p_i},r)$ with respect to $r$,
\begin{equation}
  \alpha(\vec{p_i},r) = \frac{\partial \log \rho(\vec{p_i},r)}{\partial \log r}
                      = \frac{r}{\rho}\frac{\partial}{\partial r} \rho(\vec{p_i},r) \;.
\end{equation}
In principle any differentiable kernel function and any distance measure can
be used for calculating $\alpha$. In the following we use the Euclidean norm as
distance measure and a set of Gaussian shaping function. So the expression for $\rho$
simplifies to
\begin{equation}
  \rho(\vec{p_i},r) = \sum_{j=1}^{N} e^{-(\frac{d_{ij}}{r})^q} \;,
                       d_{ij} = \| \vec{p_i} - \vec{p_j} \| \;.
\end{equation}
The exponent $q$ controls the weighting of the points according to their
distance to the point for which $\alpha$ is calculated.
In this study we calculate $\alpha$ for the case
$q=2$. Using the definition in (6) yields for the weighted scaling indices
\begin{equation}
  \alpha(\vec{p_i},r) = \frac{\sum_{j=1}^{N} q (\frac{d_{ij}}{r})^q
                                             e^{-(\frac{d_{ij}}{r})^q}}
                             {\sum_{j=1}^{N} e^{-(\frac{d_{ij}}{r})^q}} \;.
\end{equation}
Structural components of the temperature map are characterized by the calculated
value of $\alpha$ of the pixels belonging to them.
For example, points in a point-like structure have $\alpha \approx 0$
and pixels forming line-like structures have $\alpha \approx 1$.
Area-like structures
are characterized by $\alpha \approx 2$ of the pixels belonging to them.
A uniform distribution of points yields $\alpha \approx 3$ which is equal
to the dimension of the configuration space.
Pixels in the vicinity of point-like structures, lines or areas
have $\alpha > 3$.\\
The scaling indices for the whole random field under study form the frequency
distribution $N(\alpha)$
\begin{equation}
  N(\alpha) d\alpha = \#(\alpha \in [ \alpha,\alpha+d\alpha [)
\end{equation}
or equivalently the probability distribution
\begin{equation}
  P(\alpha) d\alpha = \mbox{Prob}(\alpha \in [ \alpha,\alpha+d\alpha [)
\end{equation}
This representation of the temperature map can be regarded as a
structural decomposition of the image where
the pixels are differentiated according to the local
morphological features of the structure elements to
which they belong to.\\
On the other hand, the weighted scaling indices can
be regarded as a filter response of a local nonlinear
filter acting in the CMB image.
The findings in the field of the perception and analysis image 
patterns and image textures
(for a review see Julesz 1991) suggest that, when two
textures $T_1$ and $T_2$ are discriminable,
they are distinguished by different spatial averages
of some locally computed nonlinear response function R.
Based on these considerations we calculate
the local mean values $<\alpha>$ for the
scaling indices in a sliding window of size $K$,

\begin{equation}
  <\alpha(x_i,y_i) > = \frac{1}{K} \sum_{x,y} \alpha(x_i,y_i)
                       \Theta( \frac{K}{2} - | x_i - x |)
                       \Theta( \frac{K}{2} -| y_i - y |) \;,
\end{equation}
for all pixels and analyse the respective
probability distribution
\begin{equation}
  P(<\alpha>) d\alpha = \mbox{Prob}(<\alpha> \in [ <\alpha>,<\alpha>+d<\alpha> [) \;.
\end{equation}
In all following calculation we use a window size of
$K=20$.

%%%%%%%%%%%%%%%%%%%%%%%%%%%%%%%%%%%%%%%%%%%%%%%%%%%%%%%%%%%%%%%%%%%%%%%%%
%%%%%%%%%%%%%%%%%%%%%%%%%%%%%%%%%%%%%%%%%%%%%%%%%%%%%%%%%%%%%%%%%%%%%%%%%
\subsection{Minkowski functionals}

Minkowski functionals incorporate correlation functions
of higher orders and supply global morphological information
about structures under study. It has been shown that
the $d+1$ Minkowski functionals provide a unique description
of the global morphology of a d-dimensional pattern.
In our two-dimensional temperature maps we have three
Minkowski functionals,
which can be interpreted as area ($M_0$),
circumference ($M_1$) and Euler characteristic
($M_3$) of an excursion set $R(\nu)$:

\begin{equation}
  M_0(\nu) = \int_{R(\nu)} dS  \;,
\end{equation}

\begin{equation}
  M_1(\nu) = \int_{\partial R(\nu)} dl\;,
\end{equation}

\begin{equation}
  M_2(\nu) = \int_{\partial R(\nu)} \frac{dl}{r}\;.
\end{equation}

$\partial R(\nu)$ is the boundary  of the excursion region
$R(\nu)$ at the threshold temperature $\nu$. The differentials
$dS$ and $dl$ denote the elements of area and of length along
the boundary, and $r$  is the radius of the curvature of the boundary.
The excursion set is taken as the region
of the CMB map above a certain threshold $\nu$.
The Minkowski functionals are therefore functions of the threshold
temperature $\nu$.
We estimate the Minkowski functionals for pixelized
temperature maps in flat space, which is sufficient for
our comparisons of two-dimensional random fields desired in this paper.
The calculation of the surface area and circumference
of the excursion sets in a pixelized map is straightforward.
The Euler characteristic is determined by looking at the angle deficits
of the vertices as described in Mecke (1996).
In our study, where we want to compare the original maps with
their surrogates with equal temperature distribution, the
Minkowski functional $M_0$ will - by definition - be equal for
the original and surrogate data and will therefore have no
discriminative power. Thus we will only calculate and further analyse
$M_1$ and $M_2$.

%%%%%%%%%%%%%%%%%%%%%%%%%%%%%%%%%%%%%%%%%%%%%%%%%%%%%%%%%%%%%%%%%%%%%%%%%%
%%%%%%%%%%%%%%%%%%%%%%%%%%%%%%%%%%%%%%%%%%%%%%%%%%%%%%%%%%%%%%%%%%%%%%%%%%

\section{Application to Simulated CMB Maps}
We apply the method of surrogates and weighted scaling
indices to a simulated non-Gaussian CMB map, which is a
realization of $12.8$  $\mbox{deg}^2$ CMB anisotropies due to the Kayser-Stebbin
effect from cosmic strings (e.g. Bouchet, Bennett\& Stebbins 1988).
Topological defects, although not the main source of cosmic structure,
are predicted in many particle physics models and are thought to be
responsible for the formation of cosmic strings at the end of an inflationary
period.
Fig. 1 (lower left image) shows a CMB anisotropy map generated
by cosmic strings seen after last scattering corresponding to secondary anisotropies
imprinted via the moving lens effect. Each infinitesimal element  of the cosmic string
acts as a source of 'butterfly'-pattern  whose superpositions lead
to step-like discontinuities along the string with a magnitude proportional to the
local string velocity transversal to the line of sight (LOS).
Therefore non-Gaussianities are induced by the step-like discontinuities.\\
We will test for non-Gaussianity by
generating 20 surrogate maps for the original non-Gaussian map
and for maps with additive white Gaussian noise with five different
fluctuation levels. The noise levels are chosen with $rms$ ratio
$SNR= 8,4,2,1$ and $0.5$.
In Fig. 2 the convergence of the algorithm as a function of the
iteration steps is shown. We therefore calculated the
relative deviation $\Delta I_{fourier}$ of the power spectrum
at the n-th iteration step from the original power spectrum,
\begin{equation}
  \Delta I_{fourier} = \frac{\sum_{k_x,k_y} (| I_n(k_x,k_y) | -
                                                 | I(k_x,k_y) | )^2}
                            {\sum_{k_x,k_y} | I(k_x,k_y) |^2} \;.
\end{equation}
One can clearly see that the algorithm converges quickly, reproduces
the original power spectrum with a very high accuracy  and
saturates after approximately $30$ iteration steps.
In the following all surrogates are generated using 100 iteration
steps.\\
In order to test for the Gaussianity of the obtained surrogate maps
we analysed the probability density of the Fourier phases $P(\phi)$. Therefore we
calculated the deviations $\Delta P(\phi)$,
\begin{equation}
  \Delta P(\phi) = \frac{P_{surrogate}(\phi) - <P_{random}(\phi)>}
                            {\sigma_{P_{random}(\phi)}} \;,
\end{equation}
of the surrogate maps from the probability densities $P_{random}(\phi)$
of a random phase distribution.
The mean $<P_{random}(\phi)>$ and
the standard deviation $\sigma_{P_{random}(\phi)}$ were
derived from 100 realizations of random phase distributions.
Fig. 3 shows as an example the deviations $\Delta P(\phi)$ for the case
$SNR=2$. For most of the bins $\Delta P(\phi)$ is below the
$1 \sigma$-level. Only for a few uncorrelated bins the deviation
becomes larger (smaller) than $2$ ($-2$).
Therefore, no statistically significant deviations from a random distribution
are found.
It should be noted that we did the same analysis for all other
noise levels and obtained the same results.\\
In Fig. 1 the original (noisy) images and two respective surrogates
are displayed. One might observe that the surrogate maps have 'more
granular' structures but the differences between the original and surrogate maps
are not very pronounced, especially when the noise level is increased.
In order to quantify the structures in the images we calculate the
spectrum of weighted scaling indices for all maps.
We normalized the temperature distribution so that the standard deviation
$\sigma_T$ is four at each fluctuation level. (Other normalizations are
conceivable.)  The radius for
the calculation of the weighted scaling indices $\alpha$ is chosen to
$r=1.5, 2, 3, 5$ and $6$ pixels.
The original size of the images and their surrogates is 256 $\times$ 256
pixels. In order to avoid edge effects we analyse only the inner
part (216 $\times$ 216 pixels) of the $\alpha$-
and $<\alpha>$-images further.
Likewise, we calculate the Minkowski functionals only for the inner part of the
images.\\
We start the analysis of the weighted scaling indices
by plotting the global means $<\alpha>_{global}$
as a function of the different radii (Fig. 4) for the original
map and the respective surrogates.
Recall that the differences between non-Gaussianity and
Gaussianity is reflected in the differences between the test statistics
applied to the original data
and the surrogate data. The scatter of the surrogates
leads to measures of the statistical significance of the deviation
between non-Gaussianity and
Gaussianity.
In the undisturbed case (Fig. 4 (a)) one finds that
$<\alpha>_{global}$ for the original map is
simply shifted to lower values for all radii.
This shift is due to the more 'granular-like' structure on small scales
in the surrogates.
But for all images the global mean of the scaling indices
does not change very much when the radius is increased.
Therefore, a very good discrimination between the two classes
of maps (original and surrogates) is possible for all radii.
For the noisy images (Fig. 4 (b) - (f) ) one finds larger
variations with varying radius and the original map is
only clearly discriminable from the surrogates until a
noise level of $SNR = 2$ (Fig. 4 d).\\
In Fig. 5 the global standard deviation $\sigma_{\alpha}$ of
the scaling indices as a  function of the different
radii are shown for the original map and the respective surrogates.
For this global quantity one can essentially find that for noise levels
up to $SNR = 1$ (Fig. 5 e) one can find length scales $r$ at which a
discrimination between the original and surrogates is possible.\\
Similar discrimination results can also be obtained
by calculating the global standard deviation  $\sigma_{<\alpha>}$
of the local means $<\alpha>$ (Fig. 6).
For this quantity it is remarkable that
one finds for all noise levels very different functional behavior
for the surrogates compared to the original maps. So it is very likely
that refined analyses of the slopes and/or the curvatures
of $\sigma_{<\alpha>}(r)$ will yield very good discrimination results.\\
We proceed further with a differential analyses of the scaling indices
and therefore concentrate on the probability densities $P(\alpha)$ for the
(noisy) maps and their 20 surrogates for one value of $r$ ($r=5$) (see Fig. 7).
It can clearly be seen that in the undisturbed case the differences
between the surrogates and the original map in the $P(\alpha)$ spectum
are very significant, whereas the spectral differences in the noisy cases are
not so evident in this repesentation.
In order to quantify the observed differences in the
measured probability densities $P(\alpha)$ (and
$P(<\alpha>)$ see below) we calculate the
deviation $\Delta P(\alpha)$,
\begin{equation}
  \Delta P(\alpha) = \frac{P_{original}(\alpha) - <P_{surrogate}(\alpha)>}
                            {\sigma_{P_{surrogate}(\alpha)}} \;,
\end{equation}
of the $P(\alpha)$ spectrum of the original data from the mean spectrum
as derived from the respective surrogates normalized
to their standard deviation.
Fig. 8  shows $\Delta P(\alpha)$ for all noise levels. We obtain
deviations of more than $3 \sigma$ up to a noise level of $SNR = 1$.
In Fig. 9 and Fig. 10 the probability density $P(<\alpha>)$
and deviation $\Delta P(<\alpha>)$ for the local means of the scaling
indices $<\alpha>$ are displayed. Using this quantity the differences between the
surrogates and the original maps become even more significant. The
probability distributions in the noisy cases are broader for the
surrogates (compare also Fig. 6) and shifted to higher values (no noise) or
to lower values (high noise level). Consequently  we obtain for all noise
levels systematic deviations  $\Delta P(<\alpha>)$, which have maxima
for their absolute value, which are
at least larger than $1.5 \sigma$ for the respective $<\alpha>$.
Thus a discrimination between original and surrogate maps seems possible
up to noise levels of $SNR = 0.5$.\\

We now compare these results with more standard morphological
analyses of temperature maps, which are based on Minkowski functionals.
Fig. 11 shows the circumference $M_1$ and Fig. 12 the Euler characteristic $M_2$
for all noise levels and for the original and surrogate maps.
In both cases the Minkowski functionals for the original map differ significantly
from that for the surrogate maps up to a noise level of $SNR = 1$. For
the Euler characteristic $M_2$ we can also observe that there are slight
systematic difference
between the original and surrogate maps in the case of $SNR = 0.5$.
These results are very remarkable if they are compared with those
obtained by Chiang, Naselsky \& Coles (2002).
In their study they calculate Minkowski functionals for the same CMB map
and for maps with the same power spectrum but random phases.
The authors find that an identification of non-Gaussian signatures is only
possible at most up to $SNR = 2$.
In our case where both the power spectrum and the amplitude distribution is kept
one can clearly distinguish between the original map and the surrogates even
at much higher noise levels using
Minkowski functionals. Thus we can reject the hypothesis that the CMB map
is Gaussian with same power spectrum and same intensity distribution and therefore
clearly identify non-Gaussianity at high noise levels.
The additional constraint of keeping the amplitude distribution
leads to smaller fluctuations of the Minkowski functionals for the Gaussian
surrogate maps. Therefore, a more sensitive discrimination between maps
with and without non-Gaussian signatures is possible.

\section{Conclusions}

We have shown that it is possible to generate surrogate
CMB maps which mimic both the power spectrum
{\it and} the amplitude distribution of the original
simulated map by applying the method of iteratively refined
surrogates.\\
As statistical measures being sensitive to
non-Gaussianity we calculated weighted scaling indices and Minkowski
functionals for the original and surrogate maps.
For both  measures the values for the
original data are significantly different from the values obtained
for the surrogate sets.
Therefore a clear evidence for the non-Gaussian
signatures in the maps is given.
We further tested the robustness of our approach
with respect to white Gaussian (pixel) noise and found
that a detection of the non-Gaussianity up to
SNR-ratio of 0.5 is made possible.
This result applies to both test statistics.\\
Comparing these
results with those obtained with other techniques (e.g. phase mapping)
as proposed by Chiang, Naselsky \& Coles (2002)
we find that our approach is superior concerning robustness
with respect to noise.
Due to the refined construction of the surrogate maps
we were able to detect non-Gaussian signatures using
Minkowski functionals at higher noise levels as in the
above mentioned work.\\
So both methods the construction of amplitude adjusted surrogates
and the weighted scaling indices as an estimator for local
scaling properties in image data turned out to be of great
usefulness in the detection of non-Gaussianity in CMB-maps.
Thus we obtained very promising first results
by applying new approaches derived from nonlinear sciences
to CMB map analysis.\\
It is likely that a more sophisticated
analysis of the scaling indices will even increase the
statistical significance of the discrimination results.
So the improvement of the methods proposed in this study and
their application to more realistic simulated CMB maps
incorporating mixture models
as well as to
real data obtained with the WMAP- and  Planck-satellites
will be a rich field for future studies.

\section*{Acknowledgments}

The authors want to thank Milenko Zuzic and Michael Kretschmer
for giving us the opportunity to discuss
cosmological questions in a relaxed atmosphere.

%%%%%%%%%%%%%%%%%%%%%%%%%%%%%%%%%%%%%%%%%%%%%%%%%%%%%%%%%%%%%%%%%%%%%%%%%%%%%%
\begin{figure*}
  \begin{center}
     \epsfxsize=17cm
     \epsffile{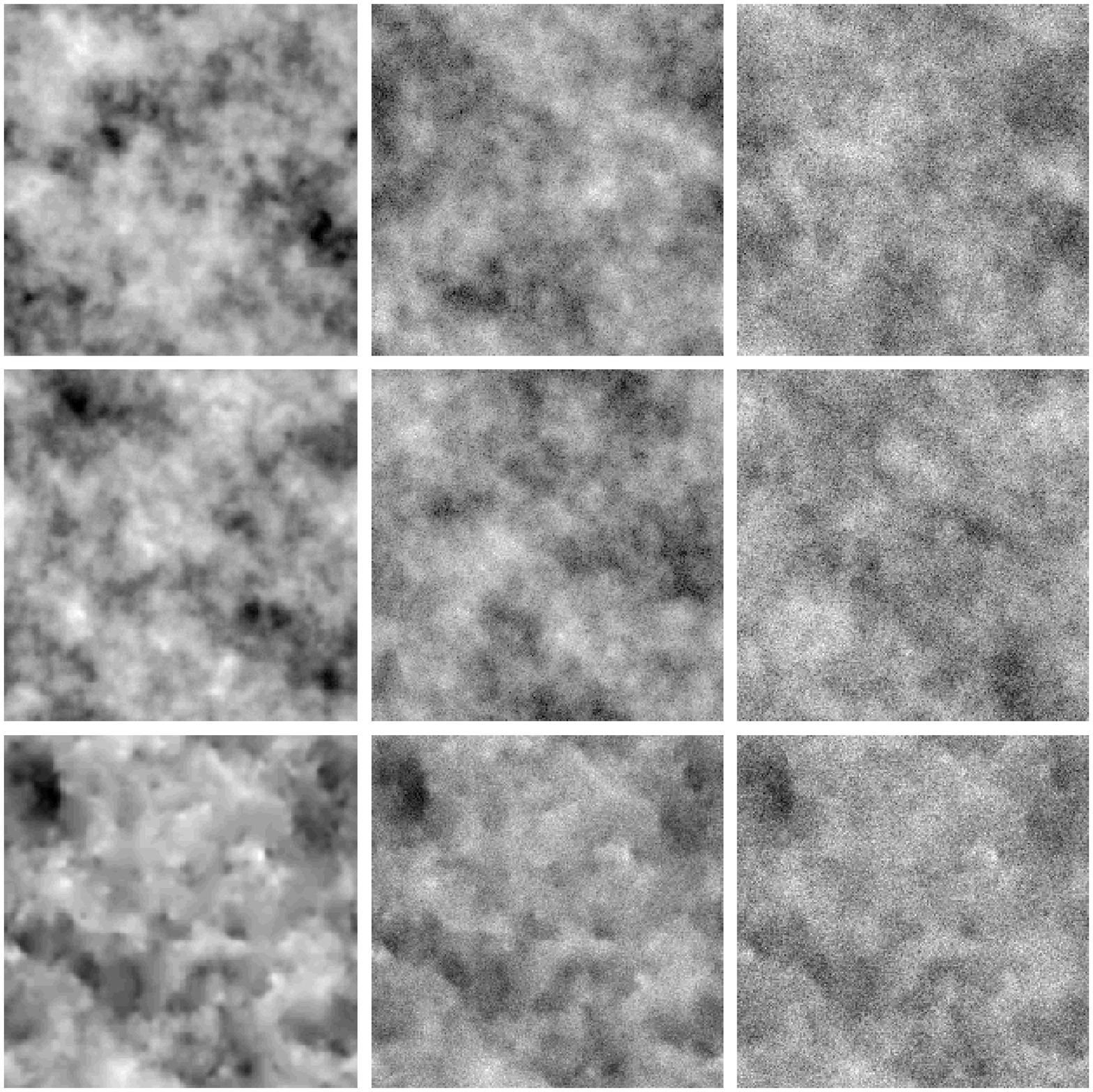}
  \end{center}
 \caption{Lowest row: Temperature map $T(x,y)$, $T \in [-100,100] \mu K$,  
          without additive noise (left image). Some step-like discontinuities
	  inducing non-Gaussianity are quite apparent.
          Temperature maps with additive noise:
          SNR = 2 (middle) and SNR = 1 (right).
	  Upper rows: Two respective surrogate images. }
\end{figure*}
%%%%%%%%%%%%%%%%%%%%%%%%%%%%%%%%%%%%%%%%%%%%%%%%%%%%%%%%%%%%%%%%%%%%%%%%%%%%%%
\begin{figure*}
  \begin{center}
     \epsfxsize=8cm
     \epsffile{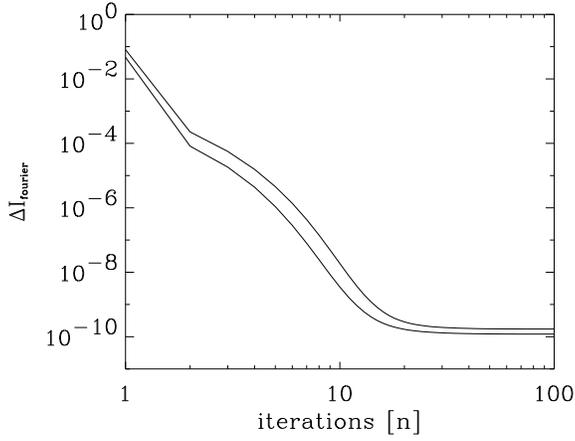}
  \end{center}
 \caption{Convergence of the iteration scheme for two noisy CMB maps with
  SNR = 4 (upper curve) and SNR=2 (lower curve). }
\end{figure*}
%%%%%%%%%%%%%%%%%%%%%%%%%%%%%%%%%%%%%%%%%%%%%%%%%%%%%%%%%%%%%%%%%%%%%%%%%%%%%%
\begin{figure*}
  \begin{center}
     \epsfxsize=8cm
     \epsffile{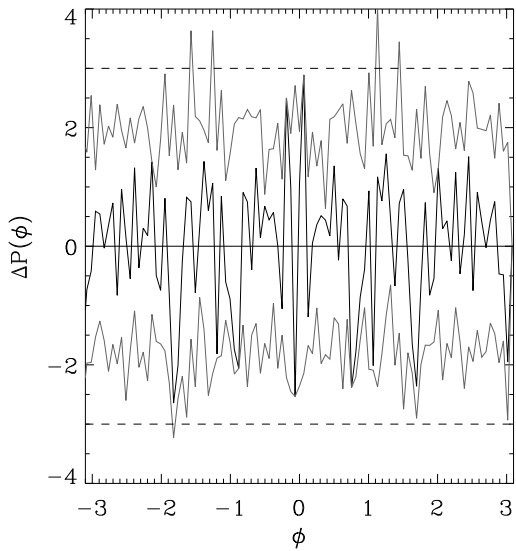}
  \end{center}
 \caption{Deviation $\Delta P(\phi)$ of the probability density of the Fourier phases
 for one surrogate $(SNR = 4)$.
  The grey lines indicate
  the maximal (minimal) deviations for each bin as derived from the $20$
  surrogates generated for this noise level.}
\end{figure*}
%%%%%%%%%%%%%%%%%%%%%%%%%%%%%%%%%%%%%%%%%%%%%%%%%%%%%%%%%%%%%%%%%%%%%%%%%%%%%%
\begin{figure*}
  \begin{center}
     \epsfxsize=17cm
     \epsffile{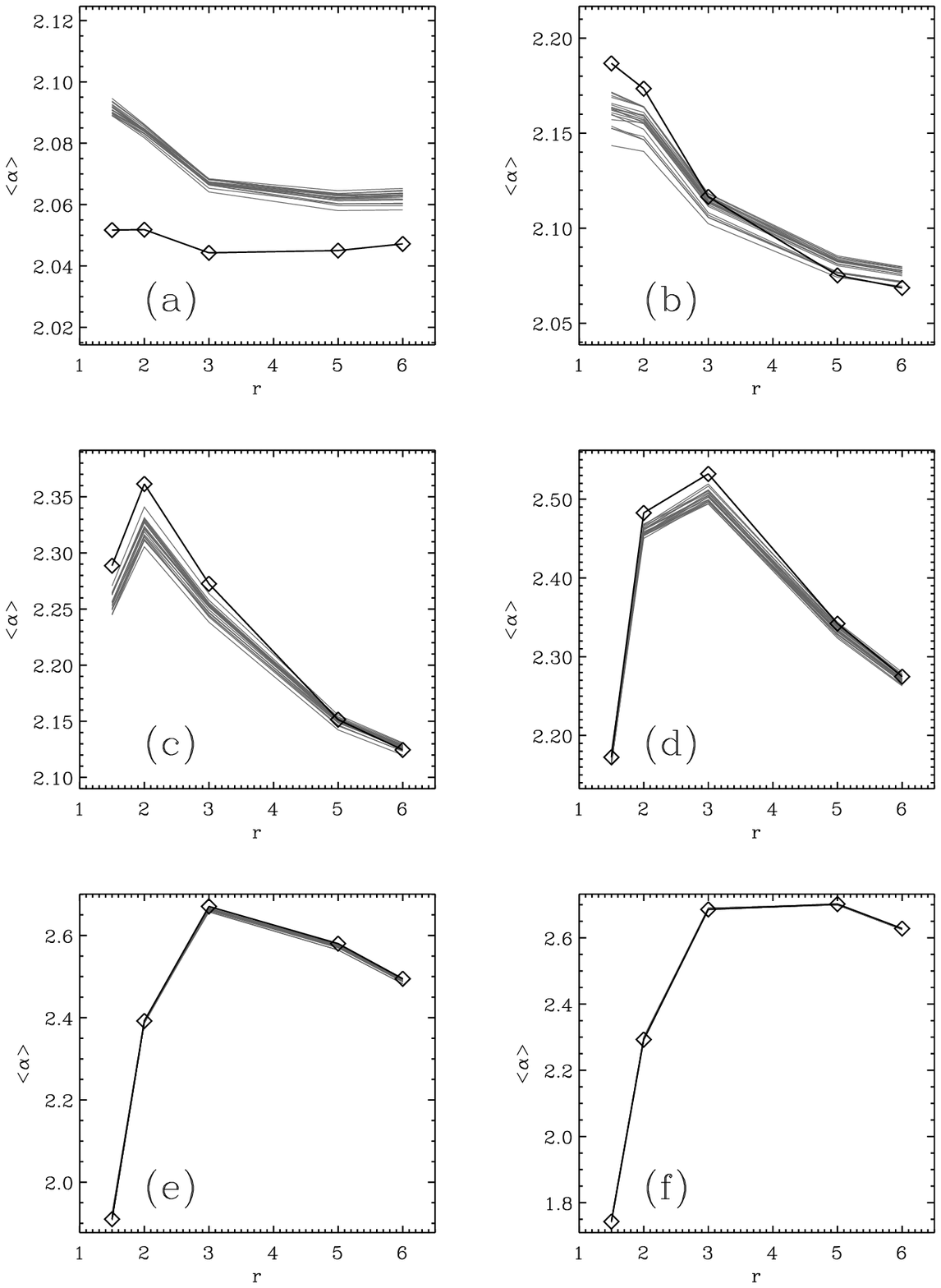}
  \end{center}
 \caption{Global mean values $<\alpha>_{global}$ as a function of $r$ for the
          original image (black) and for the surrogates (gray). Panel (a) is
	  for the pure temperature map. Panel (b), (c), (d), (e) and (f) are the
	  graphs for the combined map of CMB and additive white gaussian noise with
	  SNR = 8, 4, 2, 1 and  0.5, respectively.  }
\end{figure*}
%%%%%%%%%%%%%%%%%%%%%%%%%%%%%%%%%%%%%%%%%%%%%%%%%%%%%%%%%%%%%%%%%%%%%%%%%%%%%%
\begin{figure*}
  \begin{center}
     \epsfxsize=17cm
     \epsffile{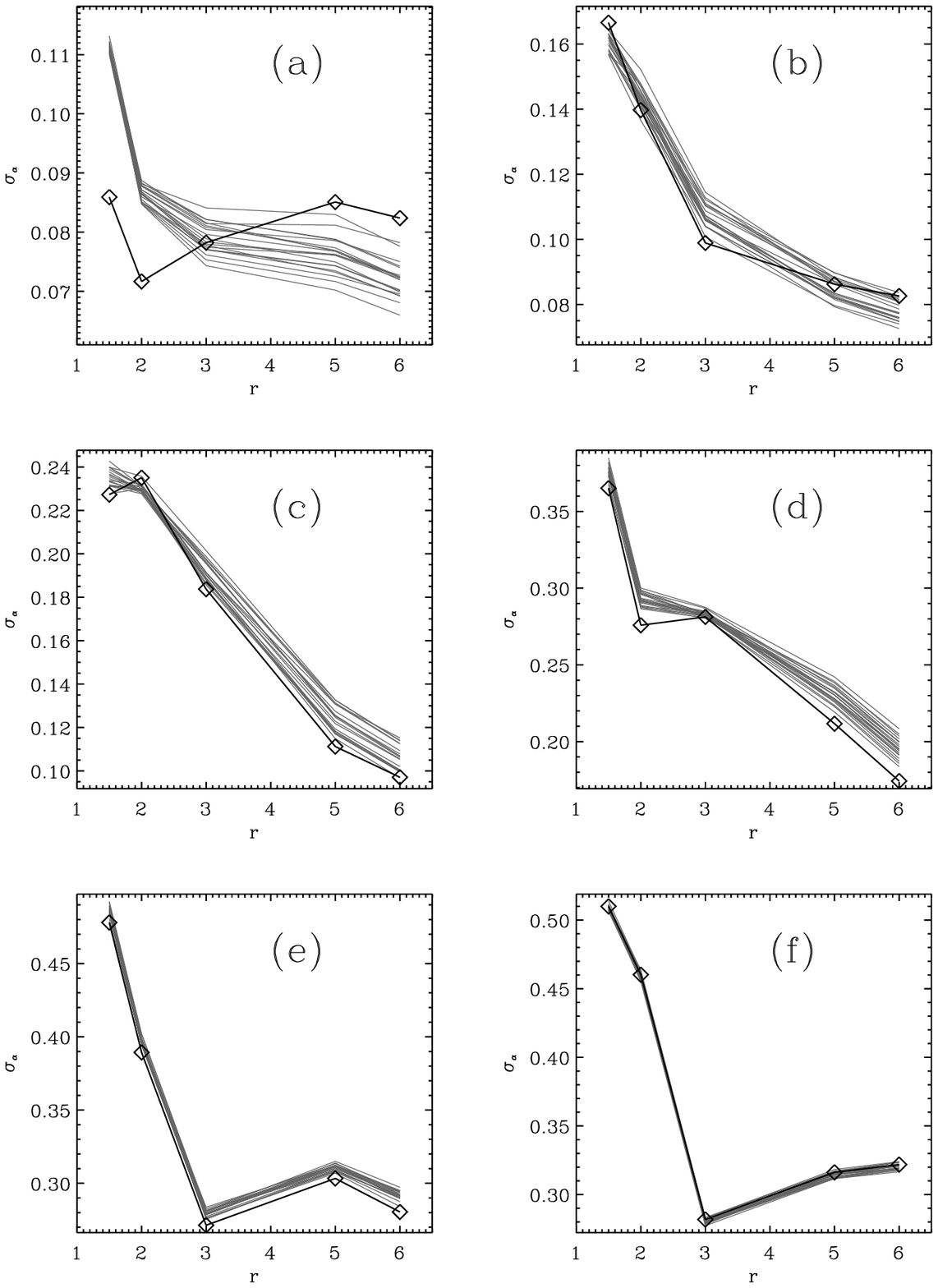}
  \end{center}
 \caption{Global standard deviation $\sigma_{\alpha}$ as a function of $r$ for the
          original image (black) and for the surrogates (gray). Panel (a) is
	  for the pure temperature map. Panel (b), (c), (d), (e) and (f) are the
	  graphs for the combined map of CMB and additive white gaussian noise with
	  SNR = 8, 4, 2, 1 and  0.5, respectively.}
\end{figure*}
%%%%%%%%%%%%%%%%%%%%%%%%%%%%%%%%%%%%%%%%%%%%%%%%%%%%%%%%%%%%%%%%%%%%%%%%%%%%%%
\begin{figure*}
  \begin{center}
     \epsfxsize=17cm
     \epsffile{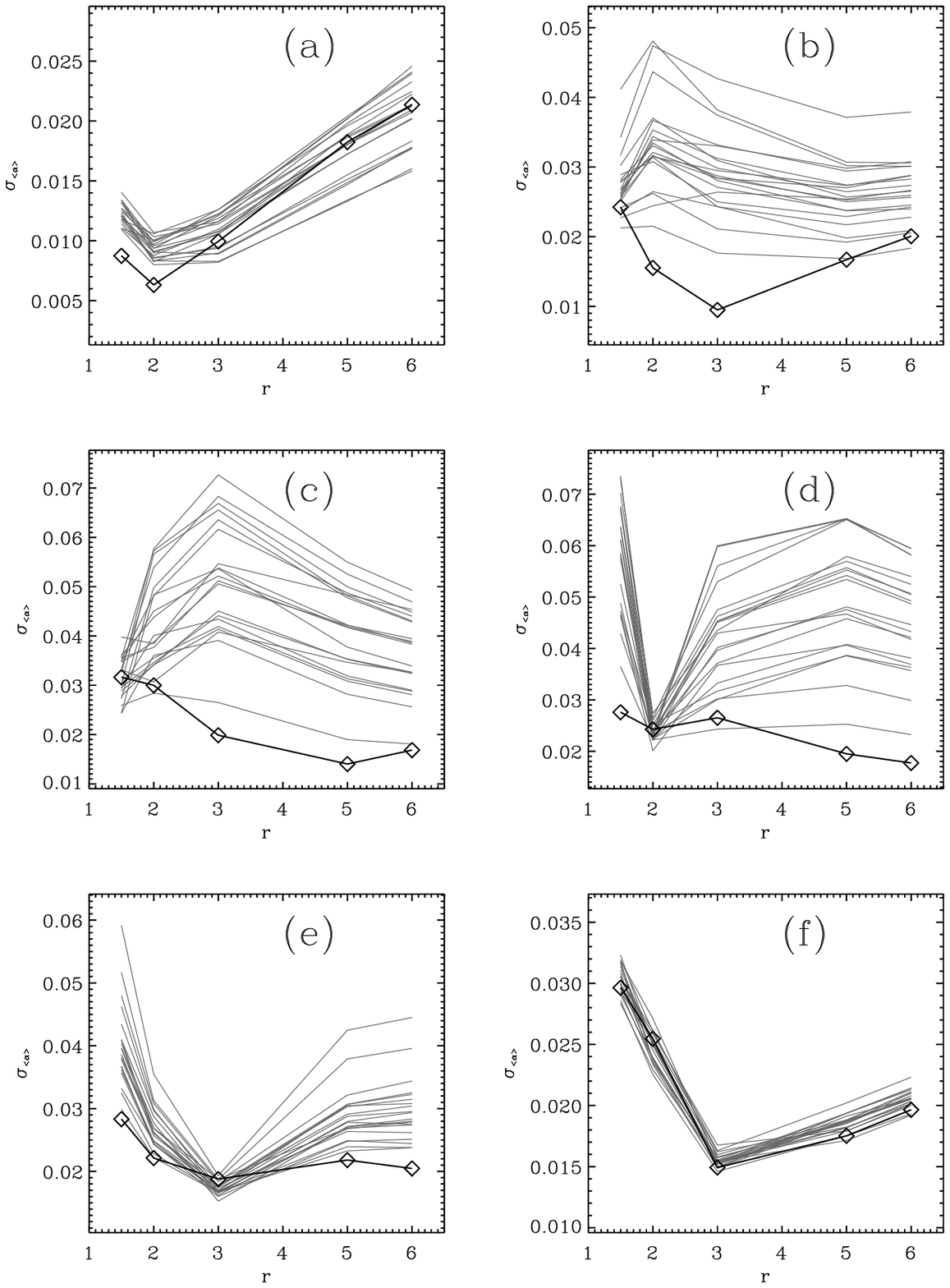}
  \end{center}
 \caption{Global standard deviation $\sigma_{<\alpha>}$ as a function of $r$ for the
          original image (black) and for the surrogates (gray). Panel (a) is
	  for the pure temperature map. Panel (b), (c), (d), (e) and (f) are the
	  graphs for the combined map of CMB and additive white gaussian noise with
	  SNR = 8, 4, 2, 1 and  0.5, respectively.}
\end{figure*}
%%%%%%%%%%%%%%%%%%%%%%%%%%%%%%%%%%%%%%%%%%%%%%%%%%%%%%%%%%%%%%%%%%%%%%%%%%%%%%

\begin{figure*}
  \begin{center}
     \epsfxsize=17cm
     \epsffile{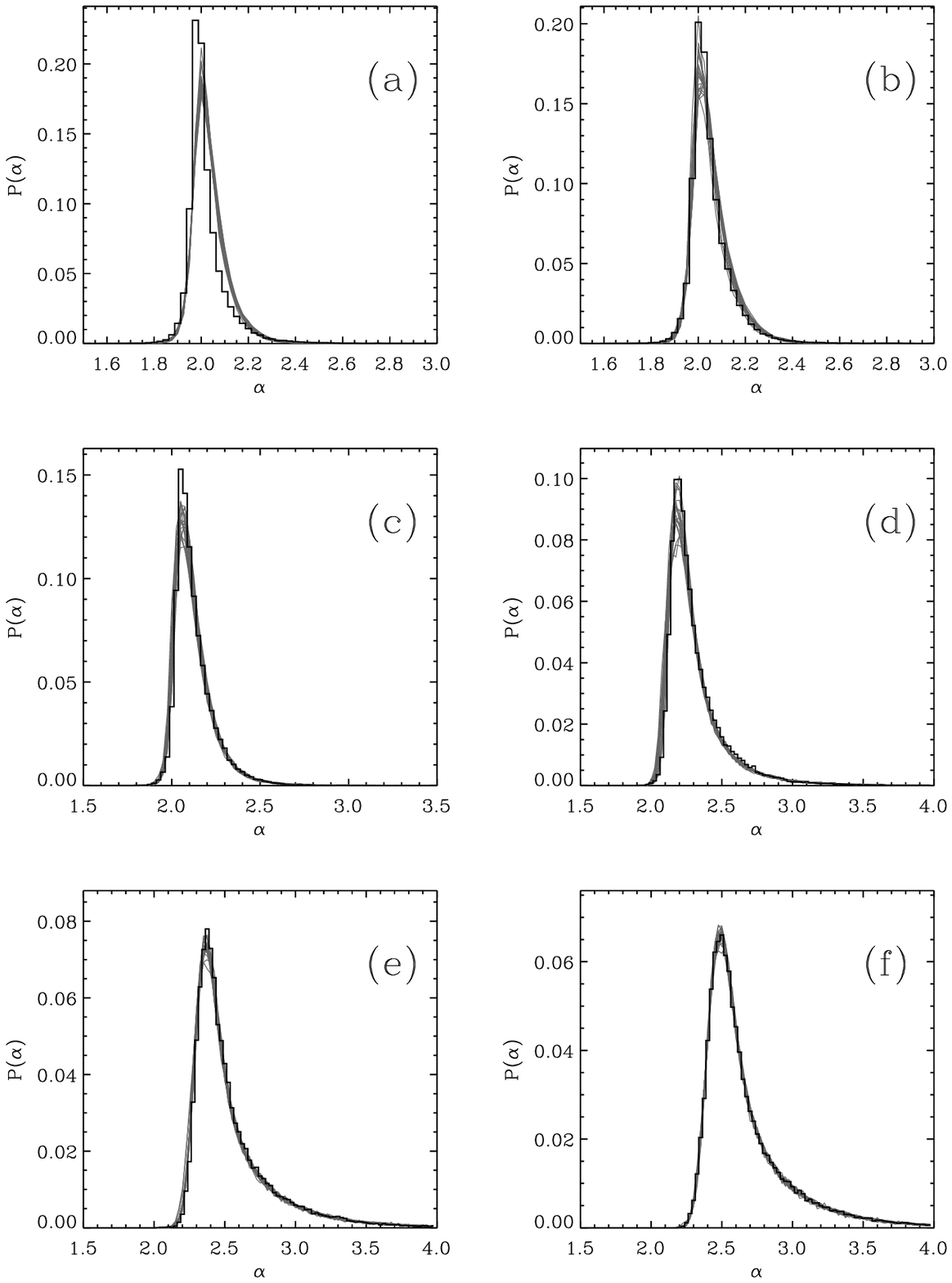}
  \end{center}
 \caption{Probability distribution $P(\alpha)$ for $r=5$ for the original
          (black) and surrogates (gray). Panel (a) is
	  for the pure temperature map. Panel (b), (c), (d), (e) and (f) are the
	  graphs for the combined map of CMB and additive white gaussian noise with
	  SNR = 8, 4, 2, 1 and  0.5, respectively.}
\end{figure*}
%%%%%%%%%%%%%%%%%%%%%%%%%%%%%%%%%%%%%%%%%%%%%%%%%%%%%%%%%%%%%%%%%%%%%%%%%%%%%%
\begin{figure*}
  \begin{center}
     \epsfxsize=17cm
     \epsffile{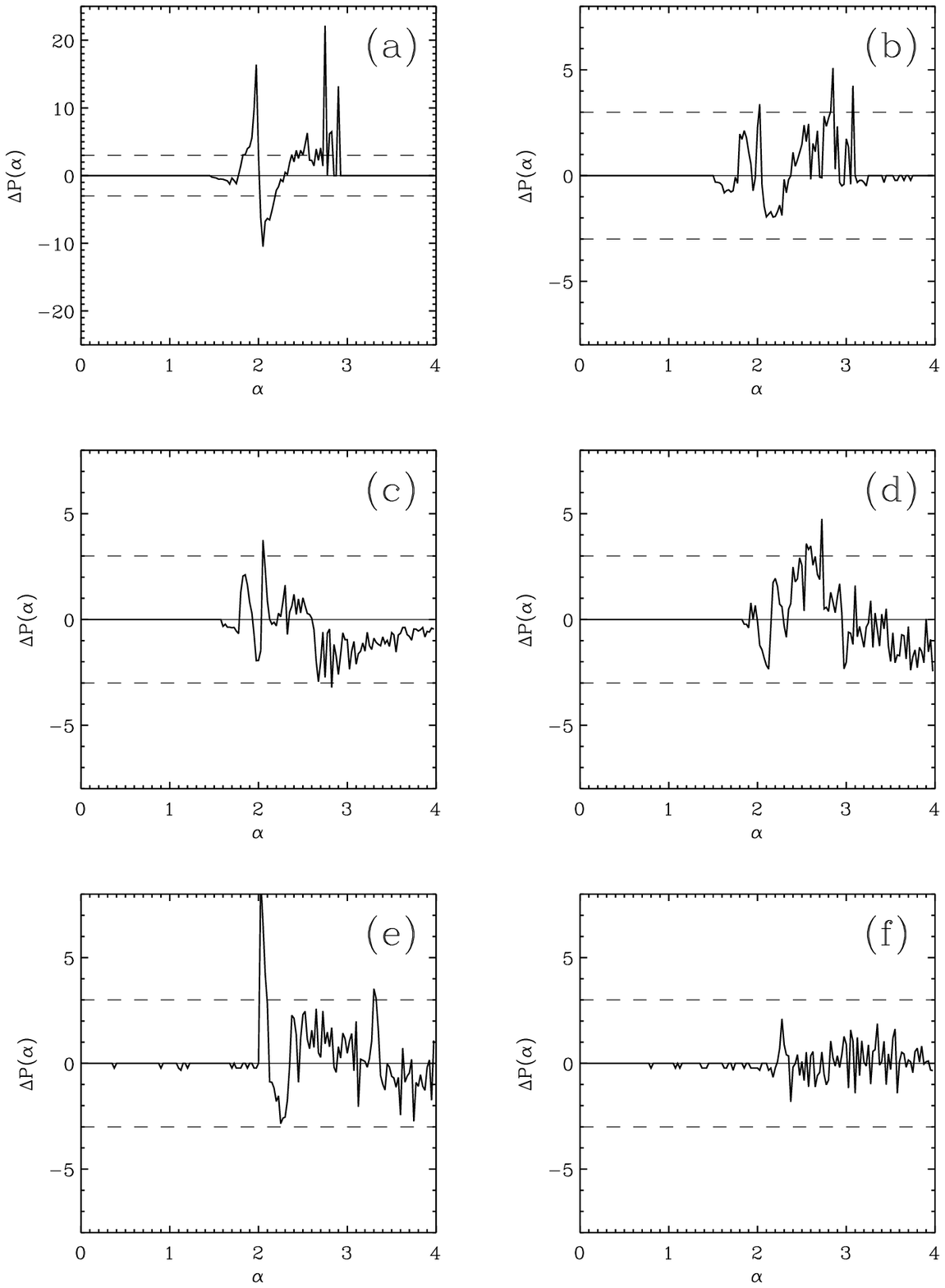}
  \end{center}
 \caption{Deviation $\Delta P(\alpha)$ of the original image from the
          mean surrogate distribution.Panel (a) is
	  for the pure temperature map. Panel (b), (c), (d), (e) and (f) are the
	  graphs for the combined map of CMB and additive white gaussian noise with
	  SNR = 8, 4, 2, 1 and  0.5, respectively.
         }
\end{figure*}
%%%%%%%%%%%%%%%%%%%%%%%%%%%%%%%%%%%%%%%%%%%%%%%%%%%%%%%%%%%%%%%%%%%%%%%%%%%%%%
\begin{figure*}
  \begin{center}
     \epsfxsize=17cm
     \epsffile{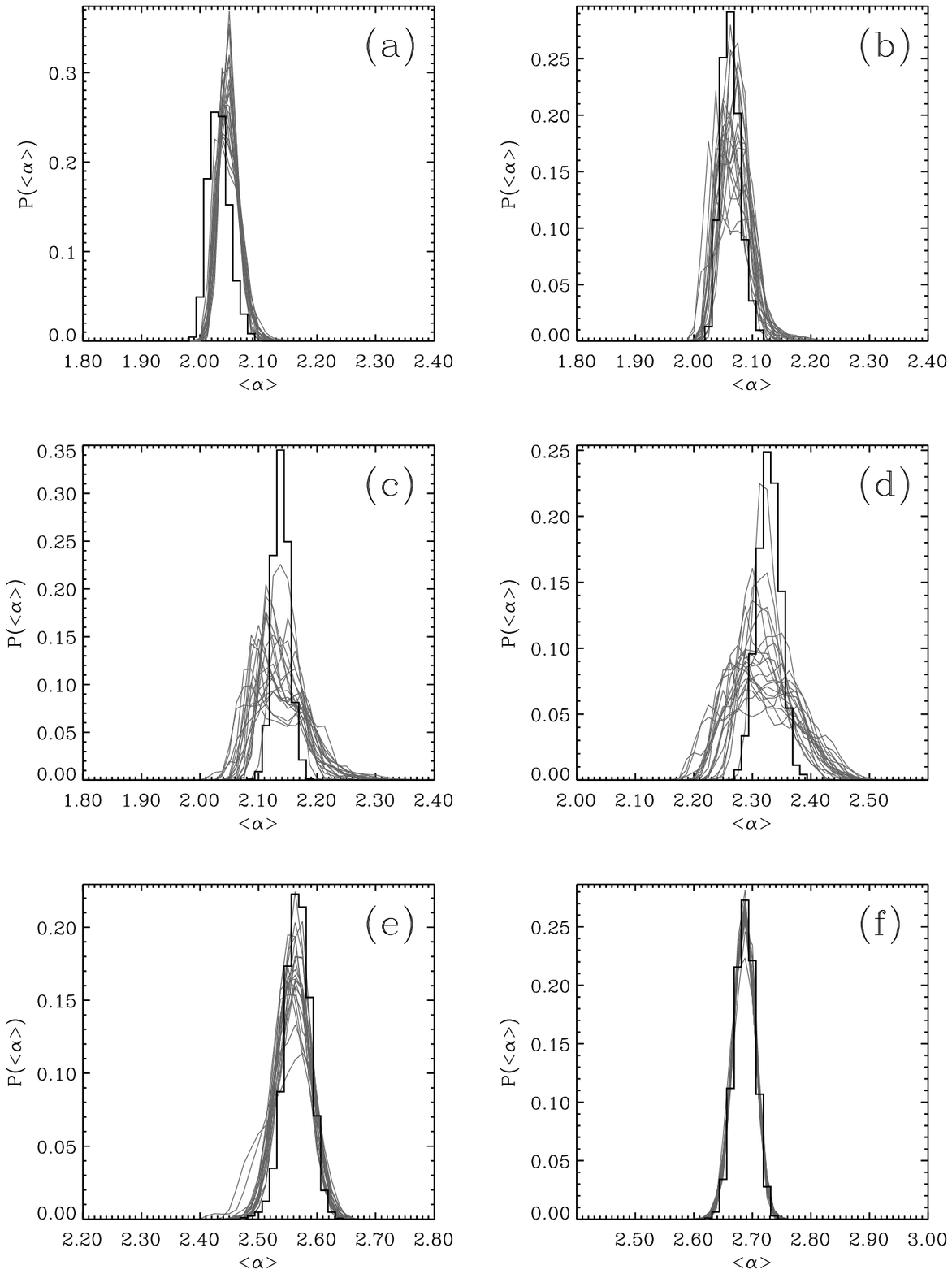}
  \end{center}
 \caption{Probability distribution $P(<\alpha>)$ for $r=5$ for the original
          (black) and surrogates (gray). Panel (a) is
	  for the pure temperature map. Panel (b), (c), (d), (e) and (f) are the
	  graphs for the combined map of CMB and additive white gaussian noise with
	  SNR = 8, 4, 2, 1 and  0.5, respectively.}
\end{figure*}
%%%%%%%%%%%%%%%%%%%%%%%%%%%%%%%%%%%%%%%%%%%%%%%%%%%%%%%%%%%%%%%%%%%%%%%%%%%%%%
\begin{figure*}
  \begin{center}
     \epsfxsize=17cm
     \epsffile{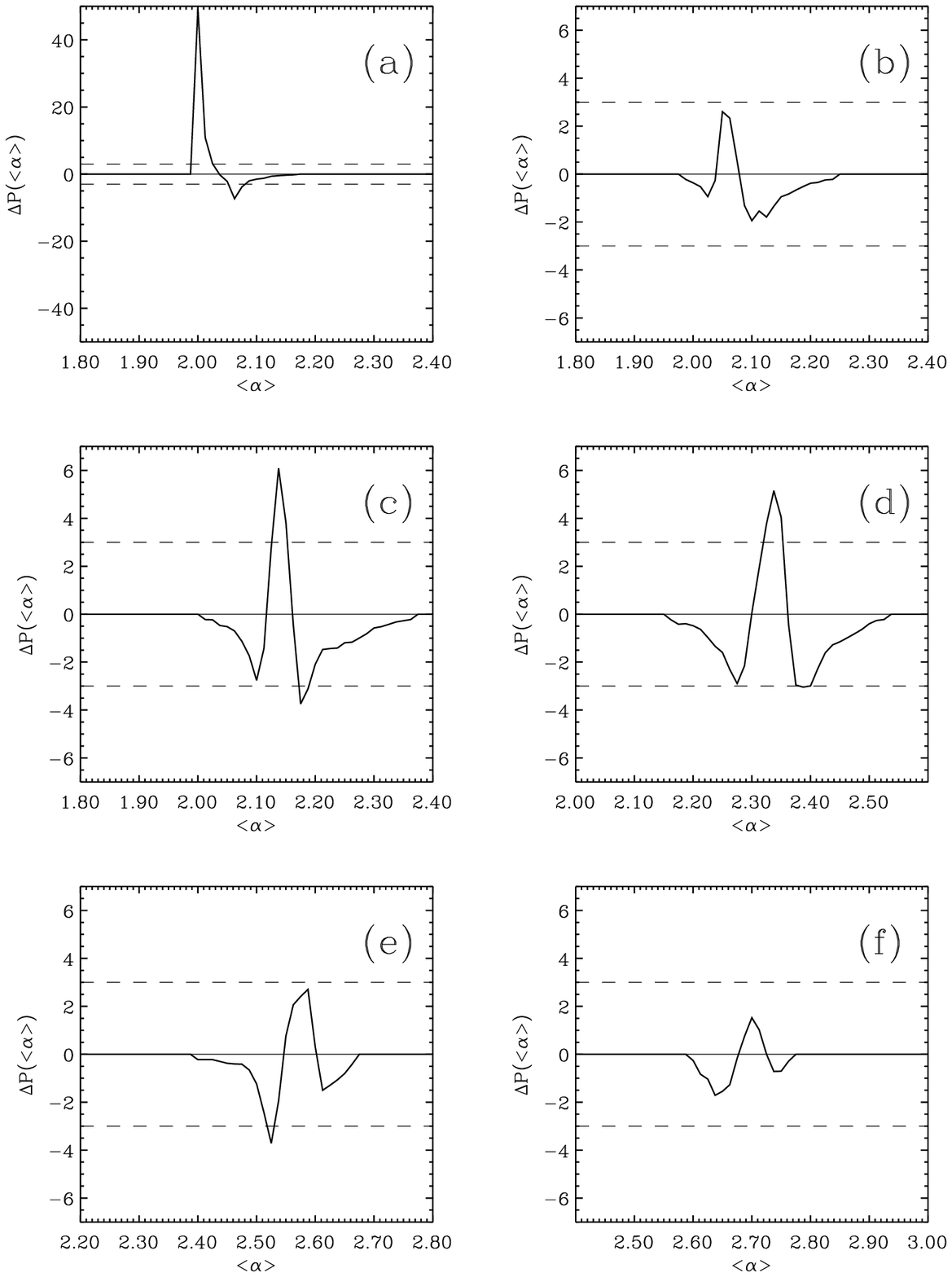}
  \end{center}
 \caption{Deviation $\Delta P(<\alpha>)$ of the original image from the
          mean surrogate distribution.Panel (a) is
	  for the pure temperature map. Panel (b), (c), (d), (e) and (f) are the
	  graphs for the combined map of CMB and additive white gaussian noise with
	  SNR = 8, 4, 2, 1 and  0.5, respectively.}
\end{figure*}
%%%%%%%%%%%%%%%%%%%%%%%%%%%%%%%%%%%%%%%%%%%%%%%%%%%%%%%%%%%%%%%%%%%%%%%%%%%%%%
\begin{figure*}
  \begin{center}
     \epsfxsize=17cm
     \epsffile{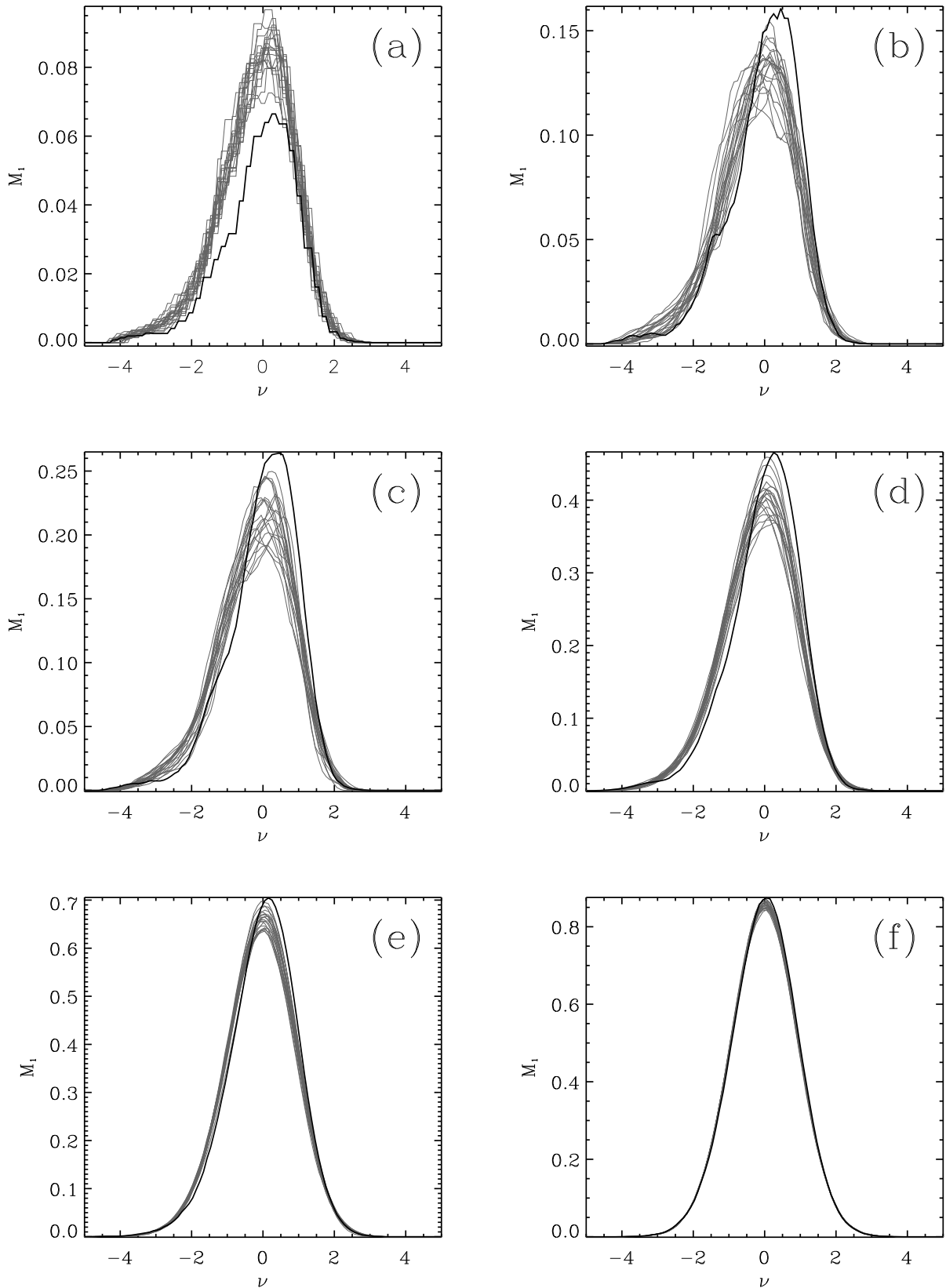}
  \end{center}
 \caption{Minkowski functional $M_1$ of the original image and the
          surrogates. The map intensities are normalized
	  to have zero mean and a standard deviation of one. The thresholds
	  $\nu$ are in units of the standard deviation. Panel (a) is
	  for the pure temperature map. Panel (b), (c), (d), (e) and (f) are the
	  graphs for the combined map of CMB and additive white gaussian noise with
	  SNR = 8, 4, 2, 1 and  0.5, respectively.}
\end{figure*}
%%%%%%%%%%%%%%%%%%%%%%%%%%%%%%%%%%%%%%%%%%%%%%%%%%%%%%%%%%%%%%%%%%%%%%%%%%%%%%
\begin{figure*}
  \begin{center}
     \epsfxsize=17cm
     \epsffile{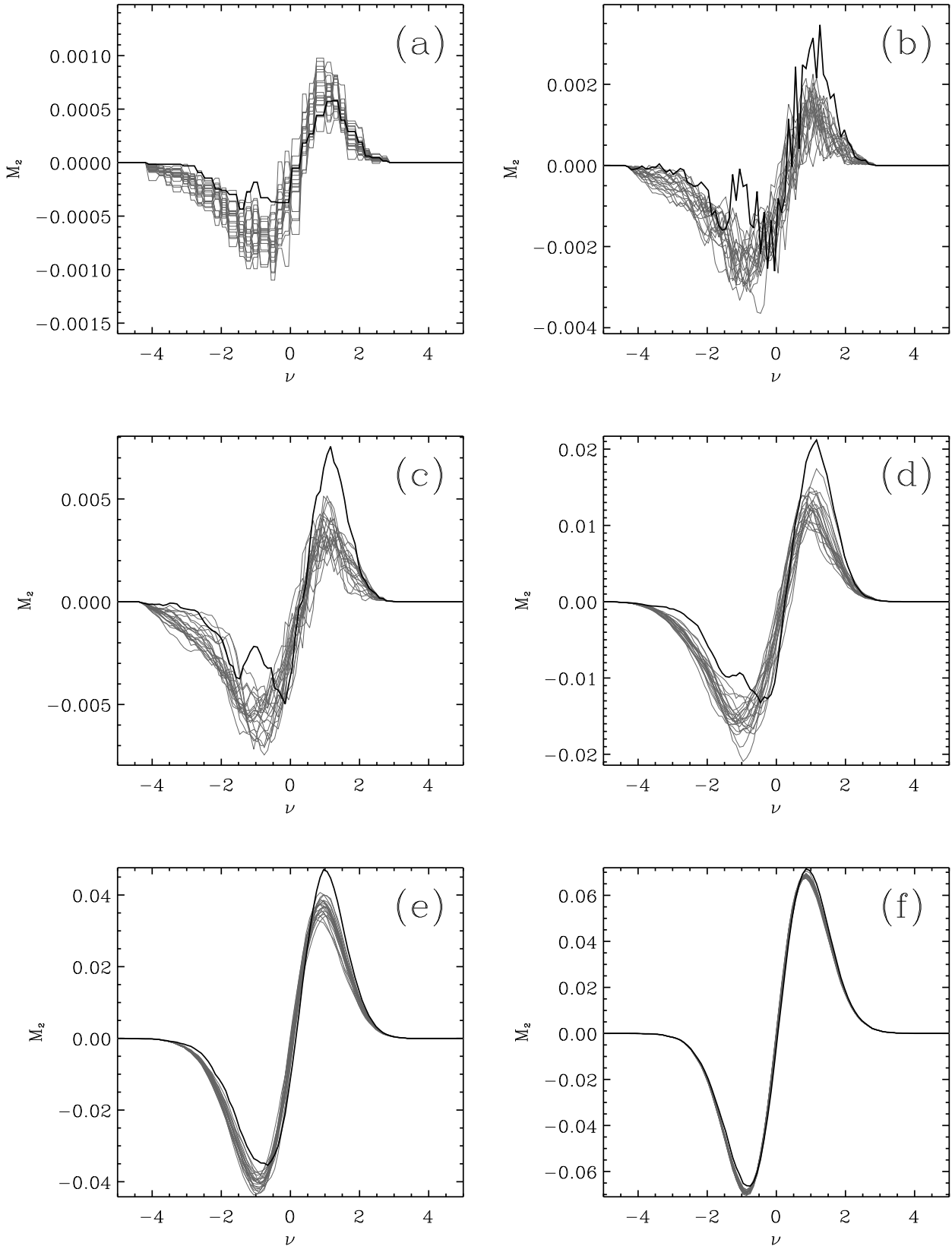}
  \end{center}
 \caption{Minkowski functional $M_2$ of the original image and the
          surrogates.Panel (a) is
	  for the pure temperature map. Panel (b), (c), (d), (e) and (f) are the
	  graphs for the combined map of CMB and additive white gaussian noise with
	  SNR = 8, 4, 2, 1 and  0.5, respectively.}
\end{figure*}

\end{document}